\documentclass[%
reprint,
superscriptaddress,
amsmath,amssymb,
aps,
prl,
floatfix]{revtex4-2}
\usepackage{amssymb}
\usepackage{amsmath}
\usepackage{graphicx}
\usepackage{siunitx}
\usepackage{comment}
\usepackage{chemformula}
\usepackage{dcolumn}
\usepackage{bm}
\usepackage{epstopdf}
\usepackage[utf8]{inputenc}
\DeclareUnicodeCharacter{0305}{\={}}

\usepackage{hyperref}
\usepackage{cleveref}

\begin{document}
\preprint{APS/123-QED}
\title{Ultrafast Proton-Induced Collective Dynamics Regulated by Water-Mediated Hydrogen-Bond Networks in Liquids}

\author{Xiaokun Liu}
\thanks{These authors contributed equally to this work.}
\affiliation{Center for Transformative Science and School of Physical Science and Technology, ShanghaiTech University, Shanghai 201210, China}
\author{Bozhu Chen}
\thanks{These authors contributed equally to this work.}

\affiliation{Songshan Lake Materials Laboratory, Dongguan 523808, China}
\author{Yizhu Zhang}
\email{zhangyizhu@tiangong.edu.cn}
\affiliation{School of Electrical Engineering, Tiangong University, Tianjin 300387, China}

\author{Jiyu Xu}
\email{jiyuxu@iphy.ac.cn}
\affiliation{Beijing National Laboratory for Condensed Matter Physics, Institute of Physics, Chinese Academy of Sciences, Beijing 100190, China}
\author{Meng Li}
\affiliation{Center for Transformative Science and School of Physical Science and Technology, ShanghaiTech University, Shanghai 201210, China}
\affiliation{Shanghai Advanced Research Institute, Chinese Academy of Sciences, Shanghai 201210, China}
\author{Yanjun Gao}
\affiliation{Shanghai Advanced Research Institute, Chinese Academy of Sciences, Shanghai 201210, China}
\author{Ji Qi}
\affiliation{School of Physics, Henan Normal University, Xinxiang 453007, China}
\author{Yuqian Yang}
\affiliation{Center for Transformative Science and School of Physical Science and Technology, ShanghaiTech University, Shanghai 201210, China}
\author{Wennan Wang}
\affiliation{Center for Transformative Science and School of Physical Science and Technology, ShanghaiTech University, Shanghai 201210, China}
\author{Sheng Meng}
\email{smeng@iphy.ac.cn}
\affiliation{Beijing National Laboratory for Condensed Matter Physics, Institute of Physics, Chinese Academy of Sciences, Beijing 100190, China}
\affiliation{Songshan Lake Materials Laboratory, Dongguan 523808, China}
\affiliation{School of Physical Sciences, University of Chinese Academy of Sciences, Beijing 100190, China}
\author{Yuhai Jiang}
\email{jiangyh3@shanghaitech.edu.cn}
\affiliation{Center for Transformative Science and School of Physical Science and Technology, ShanghaiTech University, Shanghai 201210, China}
\affiliation{School of Physics, Henan Normal University, Xinxiang 453007, China}
\date{\today}

\begin{abstract} 
Photoinduced proton dynamics in liquids couple molecular reactions to nonequilibrium charge transport, yet the role of alkyl C–H sites as proton sources remains poorly understood. Combining optical-pump terahertz-probe spectroscopy with real-time time-dependent density-functional-theory simulations, we investigate ultrafast proton dynamics following strong-field ionization of liquid ethanol and hydroxyl-free ethyl acetate. Isotope-substitution and control experiments support the identification of alkyl C–H groups as the proton source. The released protons induce quasi-free-electron localization on a timescale of hundreds of femtoseconds, manifested as a characteristic V-shaped transient in terahertz absorption. In ethanol–water mixtures, trace water enhances hydrogen-bond-network connectivity and facilitates proton transfer to ethanol O–H sites, suppressing the transient absorption dip. These findings identify photoactivated C–H sites as proton sources in liquids and reveal how hydrogen-bond networks regulate the ensuing proton and electron dynamics, linking local bond cleavage to the ultrafast charge response of the surrounding liquid.

\end{abstract}

\maketitle
Photoinduced ultrafast proton dynamics spanning from gas to condensed phases intrinsically accompany field ionization and subsequent ultrafast relaxation dynamics on a couple of ten femtoseconds to picosecond scales, resulting in the fundamental nonequilibrium physical and chemical responses \cite{Jiang2010,Wang2021PRL,Ngai2025,Loh2020,Yin2023,Worner2025}.
In aqueous solutions, proton liberation, roaming and transport governed via photoexcitation and ionization, 
solvent relaxation, proton-electron coupling and hydrogen-bond network rearrangement\cite{Garrett2005,HammesSchiffer2010}, underpins the microscopic mechanisms of biological energy transduction and photocatalysis. Proton transfer proceeds through well‑established vehicular transport\cite{Kreuer1982} and Grotthuss hopping mechanisms \cite{Agmon1995}, which have been extensively studied both experimentally and theoretically over decades. Nevertheless, most existing investigations on liquid‑phase proton dynamics have largely centered on proton transfer originating from O‑H bond  cleavage, 
 where hydroxyl groups serve as the primary proton donors and acceptors within dynamically fluctuating hydrogen‑bond networks \cite{Loh2020,Yin2023,Smith2020,Yusof2022,lin2021wolf}.

For a long time, the prevailing paradigm assumes that alkyl C‑H bonds are chemically inert with respect to free‑proton release in polar condensed‑phase surroundings. 
It is commonly accepted that C‑H bonds prefer homolytic cleavage pathways generating neutral alkyl radicals rather than bare free protons upon photoexcitation or ionization\cite{Markle2018}. While gas‑phase studies have reported proton‑release behaviors triggered by C‑H bond breaking\cite{Jiang2010,Wang2021PRL,Zhou2025NatCommun,Ngai2025}, experimental evidence directly demonstrating C‑H‑originated free‑proton generation in liquid environments remains surprisingly limited. This knowledge gap hinders our full understanding of C‑H bond activation, which lies at the heart of catalytic functionalization of hydrocarbon molecules.

Advances in ultrafast spectroscopy have opened opportunities to track transient proton‑related dynamics in real‑time. X‑ray absorption spectroscopy has been extended to liquid samples and enables site‑selective probing of proton‑transfer dynamics with element specificity \cite{Smith2020,Loh2020,Yin2023,Worner2025}. Even so, it remains experimentally challenging to capture the   collective electronic response
accompanying proton generation, especially the transient evolution of quasi‑free electrons strongly coupled to proton motion within fluctuating hydrogen‑bond networks. Complementary experimental probes that are simultaneously sensitive to collective solvent motion and transient  quasi-free electron
populations are highly desired. 


Real-time tracking  of these dynamics requires techniques sensitive to the hydrogen-bond network in both structural and electronic degrees of freedom \cite{Brunig2022,Pezzotti2025}. 
Time‑resolved terahertz (THz) absorption spectroscopy represented by optical‑pump terahertz‑probe (OPTP) measurements
\cite{li2025apl}
offers such a unique capability. THz radiation couples strongly to low‑frequency collective motions of solvent networks
\cite{Savolainen2013} 
and is highly sensitive to the density of quasi‑free electrons produced after strong‑field ionization \cite{Knoesel2001,Savolainen2014,Runge2023,Runge2025,Novelli2023}. 
Since THz absorption is directly correlated with macroscopic transient conductivity of liquid solutions \cite{Knoesel2001,Knoesel2004,Tielrooij2009, Novelli2025}, it can report on electron localization‑delocalization dynamics induced by charged species such as newly liberated protons, providing distinctive information inaccessible from optical or X‑ray techniques alone.

Upon strong-field photoionization, the system undergoes a complex ultrafast cascade of non-equilibrium dynamics: 
an initial electron ejection on the sub-femtosecond (\SI{<1}{fs}) scale is followed by the evolution of primary cations through charge or proton transfer with neighboring molecules within hundreds of femtoseconds  \cite{Yin2023,Loh2020,Savolainen2014}. 
Subsequently, the ejected electrons are captured by the solvent environment to form "solvated electrons" on a sub-picosecond (\SI{<1}{ps}) scale \cite{Savolainen2014,Knoesel2004,Ghalgaoui2021,Singh2022,Runge2023,Runge2025,Novelli2025,Garrett2005}. The collective dynamics of quasi-free electrons in solution give rise to coherent motions at THz frequencies, 
thereby leading to characteristic absorption features in the THz regime \cite{Ghalgaoui2021,Singh2022,Runge2023}. 
During this transient regime, the liberation and transfer of protons modulate the surrounding solvated electrons, which in turn manifests as measurable changes in THz absorption. 
 In this Letter,  we carry out OPTP experiments on liquid ethanol and hydroxyl‑free ethyl acetate, combined with real‑time time‑dependent density‑functional‑theory (rt‑TDDFT) simulations. We observe characteristic V‑shaped modulations in terahertz absorption originating from transient quasi‑free‑electron localization driven by protons released from C‑H bond cleavage. Isotope substitution and control experiments unambiguously identify alkyl C‑H rather than O‑H groups as the proton donors. We further demonstrate that trace water can reconstruct fragmented hydrogen‑bond networks and regulate C‑H activation. 


 
\begin{figure}[htbp]
    \centering
    \includegraphics[width=\linewidth]{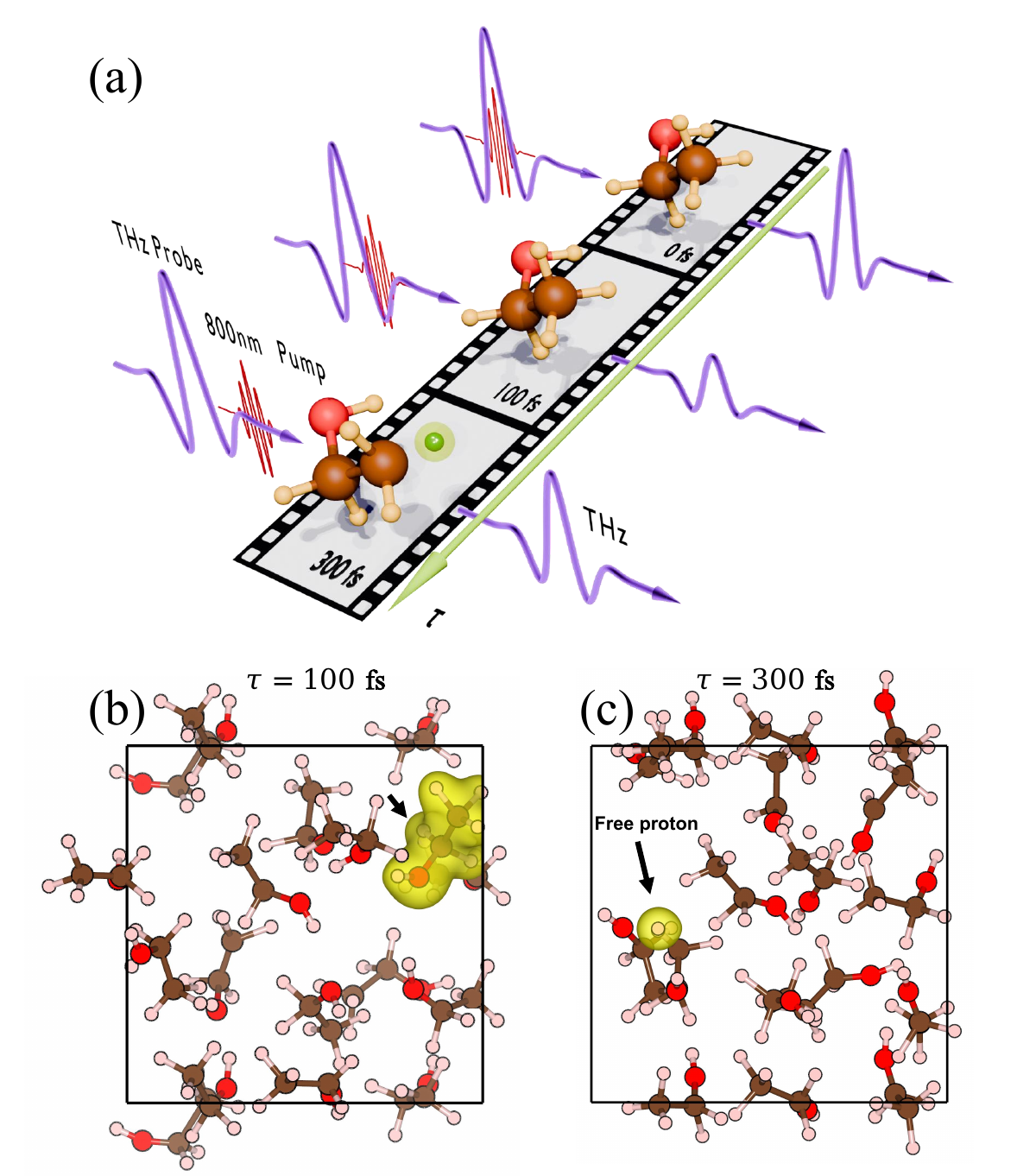} 
    \caption{ Schematic illustration of the experiment and theoretical simulation of electron density. (a) The schematic is structured into three parts: 800 nm and THz incident lights (left), ethanol sample (middle), and THz transmitted light (right). 
     Brown, red, and pale yellow spheres in ethanol sample represent carbon, oxygen, and hydrogen atoms, respectively, while the green sphere represents a free proton. 
    (b) and (c) Theoretically simulated electron density distributions around one ethanol molecule and one free proton at delay times of \SI{100}{fs} and \SI{300}{fs}, respectively. 
    }\label{fig1}
\end{figure}

The experimental schematic is shown in Figure~\ref{fig1}(a): an \SI{800}{nm} femtosecond pulse is used to induce ionization of ethanol molecules, while a THz 
pulse probes the concomitant changes in the solution's conductivity in real time. On the experiment, a \qty{800}{nm} laser beam with the pulse width \qty{60}{fs} was split into three beams for pumping the solution (\qty{0.7}{mJ}), the generation of THz (\qty{1.7}{mJ}) and the electro-optic sampling (EOS) of the THz time-domain waveform\cite{Zhao2024cp,Zhang2020ol}, respectively.  The pump beam was focused onto the solution sample 
with the focus of \SI{50}{\micro\meter} in diameter, delivering the intensities in the range of \SIrange{3e13}{11e13}{\watt\per\centi\meter\squared}. 
THz 
pulses with \qty{60}{\micro\watt} were generated via two-color strong-field ionization 
of air
and the spectral bandwidth responsible to the EOS is given by \qtyrange{0.1}{3}{THz}. With the focus of \qty{300}{\um} in diameter, the THz beam was focused onto  circulating liquid film  of an $\sim$\qty{50}{\um} thick by an off-axis parabolic mirror, spatially and temporally overlapping with the pump beam. Theoretical simulations show the electron density distribution around one ethanol molecule at delay time 100 fs and the emergence of free protons at 300 fs following laser irradiation in Fig.\ref{fig1}(b) and (c), respectively. 
A detailed description of the optical setup and liquid film generation can be found in the Supplementary Material.
\begin{figure*}[!htb]
   \centering
    \includegraphics[width=\textwidth]{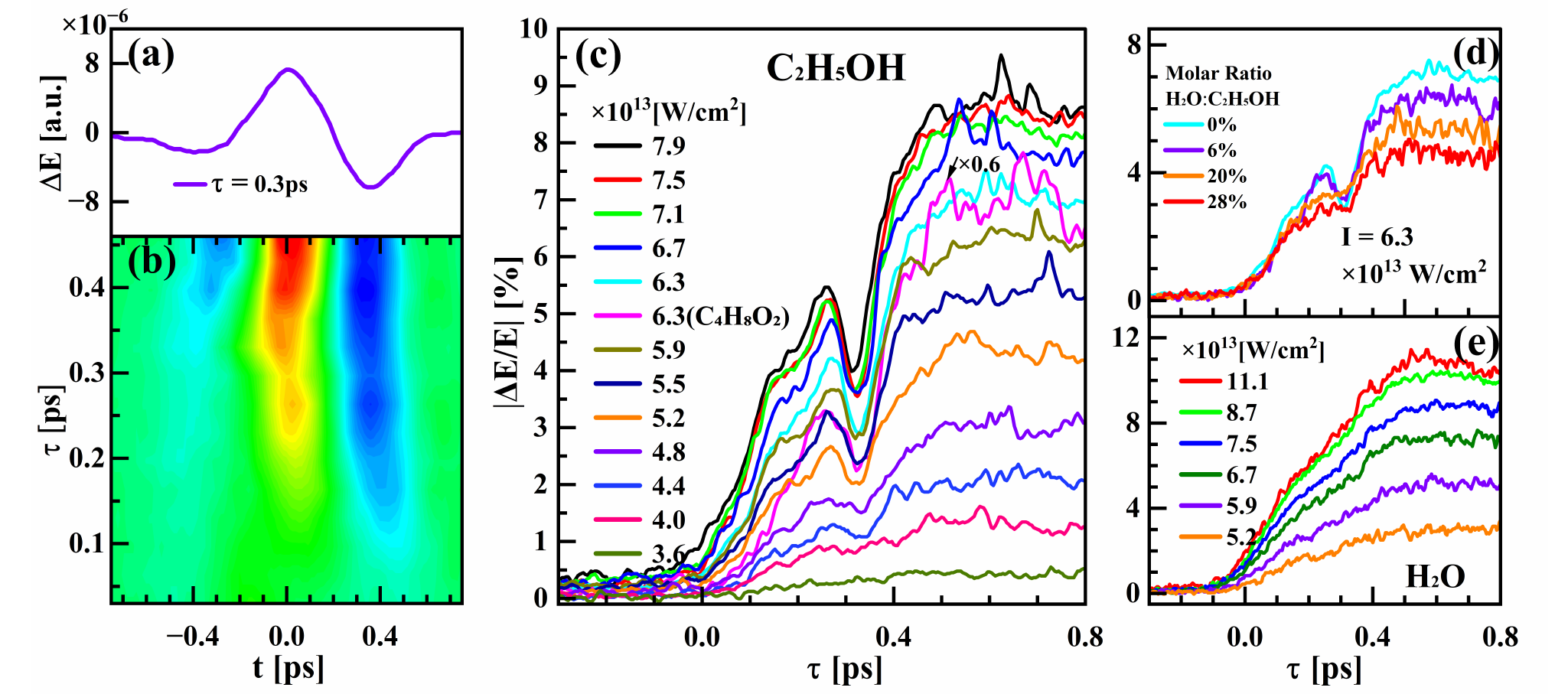} 
    \caption{Experimental optical-pump THz-probe spectroscopy.
    (a) Transient THz waveform slices $\Delta E(t)$ at selected pump--probe delays $\tau \approx$ 0.3 ps.
     (b) Two--dimensional evolution of the pump-induced change in the THz electric field, $\Delta E(\tau,t)$,
    as a function of THz sampling time $t$ and pump-probe delay $\tau$. 
    (c) Evolution of the THz absorption change \(|\Delta E/E|\)  extracted at $t \approx 0$~ps in (b) as a function of pump--probe delay $\tau$ for pure ethanol at varying pump intensities. 
    The pink curve represents the ethyl acetate trace measured at $I = 6.3\times10^{13}$~\unit{\watt\per\centi\meter\squared}, scaled by a factor of 0.6. 
    (d) \(|\Delta E/E|\) as a function of $\tau$ for ethanol--water mixture with water mole fractions of 0\%, 6\%, 20\%, and 28\% at a pump intensity $I = 6.3\times10^{13}$~\unit{\watt\per\centi\meter\squared}.
    (e) The same as (c), but for pure liquid water. 
    }\label{fig2}
\end{figure*}

To track the photoinduced ultrafast proton dynamics, 
we employ real-time time-dependent density functional theory (rt-TDDFT) simulations\cite{Lian2020} for neat ethanol and ethanol–water mixture solutions. 
The solution system was first equilibrated for 5 ps at 300 K in the constant particle number-volume-temperature (NVT) ensemble, and the last atomic configurations were irradiated by laser pulses with a gaussian envelope, $E(t) = E_0 \exp\left[-\frac{(t - t_0)^2}{2\sigma^2}\right] \cos(\omega t)$.
We use the laser pulses with a wavelength of 800 nm, a width of $\sigma = 4$ fs centered at $t_0 = 20$ fs, denoting the peak time of the pump pulse.
The photoinduced atomic dynamics are tracked up to 500 fs, and the electronic THz 
transmissivity is calculated based on electronic conductivity. More details of simulation are discussed in the Supplementary Material. 

A representative transient THz waveform $\Delta E(t)$ at specific delay, $\tau \approx 300$ fs, is plotted
in Fig.~\ref{fig2}(a) to illustrate the field attenuation. Here, $\Delta E(t)$ is the pump-induced change in the transmitted THz electric field. 
The two-dimensional evolution map of the photo-induced THz field change, $\Delta E(t, \tau)$, is presented in Fig.~\ref{fig2}(b). 
Fig.~\ref{fig2}(c) shows the evolution of $|\Delta E/E|$ in THz transmission at $t =0$~ps as a function of the pump-probe delay $\tau$ at different pump intensities $I$. Here, $E$ denotes the peak electric-field amplitude of the THz waveform transmitted through the liquid film in the absence of the pump beam, obtained from the corresponding THz time-domain waveform.
The transient THz absorption change originates from liquid conductivity dynamics, governed by $\vert{}\frac{\Delta E(\tau)}{E}\vert{} \propto \Delta \sigma(\tau) \propto \Delta n(\tau)$, where the conductivity $\sigma(\tau)$ scales directly with the charge carrier concentration $n(\tau)$. Photoinduced quasi-free electrons act as the dominant species driving this conductivity variation, thereby modulating the THz absorption.

Fig.~\ref{fig2}(e) shows the OPTP results of typical ionized aqueous solutions such as water.
The pump beam ionizes the solution, generating quasi-free electrons. These quasi-free electrons absorb THz radiation, leading to a decrease in the THz electric field transmitted through the liquid, hence $\Delta E < 0$. For $0 < \tau < 500$\,fs, the rapid increase in $|\Delta E(\tau)/E|$ indicates a rapid increase in the density of quasi-free electrons in the liquid due to photoionization. The photoionization process itself occurs within \SI{60}{fs}, but the rise in quasi-free electron density lasts for about \SI{500}{fs}, which is attributed to the solvation process of free electrons in the solution~\cite{Savolainen2014,Knoesel2004,Knoesel2001}. For $\tau > \SI{500}{fs}$, $|\Delta E(\tau)/E|$ slowly decreases, indicating a slow reduction in the quasi-free electron density due to the recombination with solution cations. 

However, in Fig.~\ref{fig2}(c), the $|\Delta E(\tau)/E|$ curve for the ethanol solution exhibits a clear difference compared to experimental results for  water in Fig.~\ref{fig2}(e). In addition to the rapid rise and slow relaxation processes similar to those observed in water, the ethanol solution's $|\Delta E(\tau)/E|$ shows an ultrafast V-shaped dip during the rise. This structure manifests as a sudden drop in THz absorption near $\tau \approx 250-350$ fs, followed by a quick recovery to the original level. The non-monotonic evolution is attributed to a sudden decrease and subsequent restoration in the electron density.

Notably, we measured the evolution of the THz absorption in ethanol–water mixtures (Fig.~\ref{fig2}(d)), where the  V-shaped dip is highly sensitive to water content and vanishes at water molar ratios up to 20\%. The gradual attenuation of this V-shaped THz dip with increasing water concentration points to enhanced free proton recombination via water-mediated hydrogen-bond networks, as elaborated in the theoretical analysis below.

\begin{figure}[htbp]
    \centering
    \includegraphics[width=\linewidth]{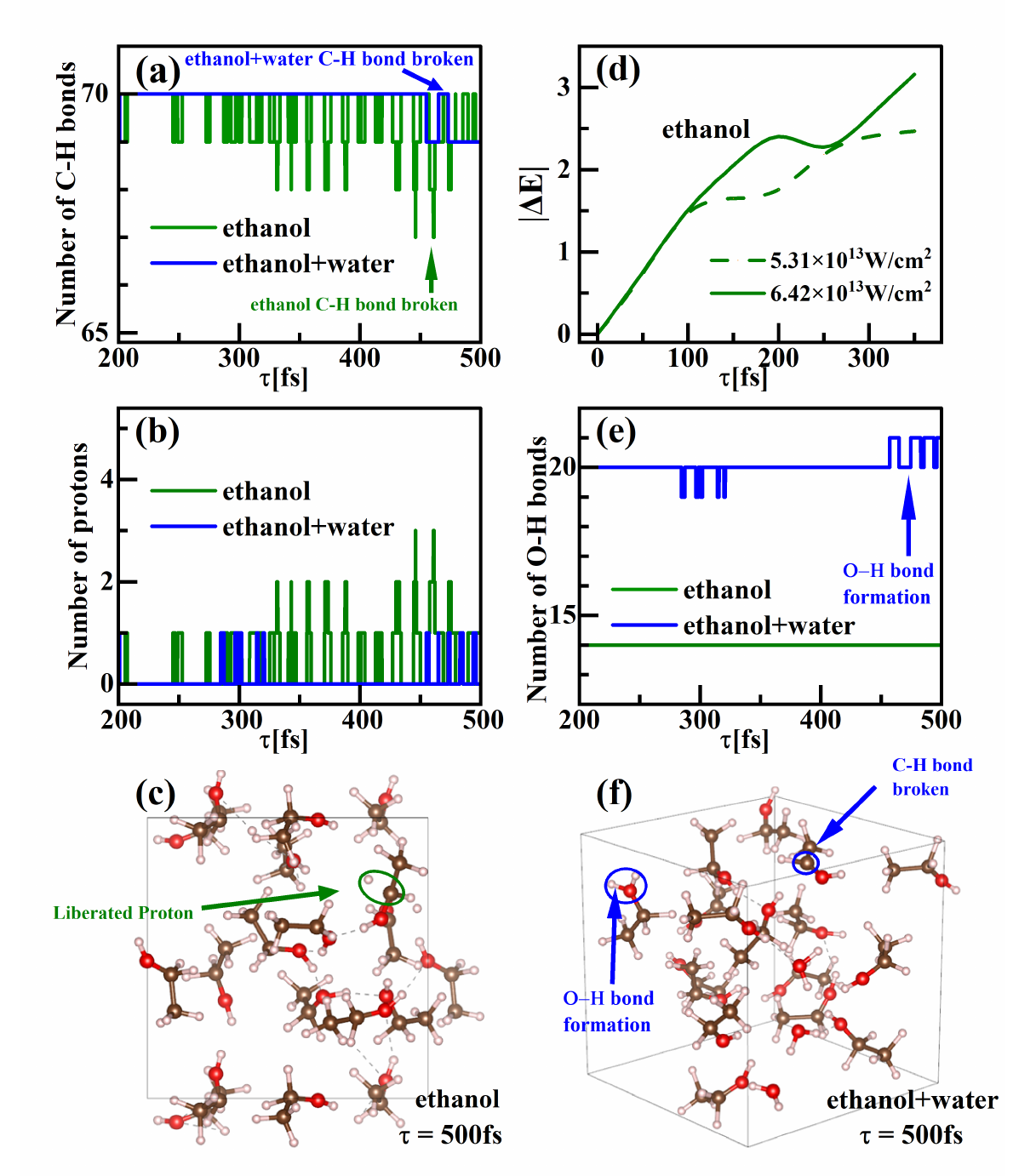}
    \caption{Simulation results from rt-TDDFT. (a) Number of C--H bonds, \(N_{\mathrm{C-H}}\), within the unit cell for pure ethanol (green) and ethanol-water mixture (blue). 
    (b) Statistics of the number of \ch{H^+}, $N_{\mathrm{H^+}}$. 
    (c) Schematic illustration of the ethanol unit cell at $\tau = \SI{500}{fs}$, showing C--H bond cleavage and a free proton. 
    (d) Theoretically simulated change in THz absorption, $\Delta E/E$, as a function of pump-probe delay $\tau$ under different pump intensities. (e) Statistics of the number of O--H bonds, $N_{\mathrm{O-H}}$. The blue arrow indicates the increase at $\tau = \SI{500}{fs}$. (f) Schematic illustration of the unit cell for the ethanol-water mixture at $\tau = \SI{500}{fs}$, depicting C--H bond cleavage and the process of bonding with $\ch{OH_2}$.}
    \label{fig4}
\end{figure}


To clarify the origin of the sudden decrease in electron density, we performed simulations using rt-TDDFT. Fig.~\ref{fig4} presents the statistical results of chemical bonds after strong-field photoionization. The curves in Figs.~\ref{fig4}(a) and \ref{fig4}(b) represent the time-dependent statistics of the number of C--H bonds, \(N_{\mathrm{C-H}}\), and the number of free protons, $N_{\mathrm{H^+}}$, 
in the ethanol system after photoionization, respectively. The criterion for identifying a free proton is that the interatomic distance exceeds the bonding distance by 20\%. For $\tau < 200$ fs, \(N_{\mathrm{C-H}}\) and $N_{\mathrm{H^+}}$ lack clear statistical significance due to chemical bond vibrations. In the interval $200$ fs $< \tau < 500$ fs, the average values of \(N_{\mathrm{C-H}}\) and $N_{\mathrm{H^+}}$ over this period reflect the numbers of C--H bonds and free protons in the ethanol system. The theoretical simulations reveal that after strong-field ionization
of liquid ethanol, \(N_{\mathrm{C-H}}\) decreases while $N_{\mathrm{H^+}}$ increases. This indicates that the cleavage of C--H bonds leads to proton dynamics, 
forming free protons in the solution. Figure~\ref{fig4}(c) shows a schematic illustration of C--H bond breaking in the ethanol unit cell at $\tau = 500$ fs, corresponding to the trends of \(N_{\mathrm{C-H}}\) and $N_{\mathrm{H^+}}$ observed in Figs.~\ref{fig4}(a) and \ref{fig4}(b).

As shown in Fig.~\ref{fig1}(b), the electron density is localized around the molecules, and the C-H bonds remain intact according to the simulations. At $\tau = \SI{300}{fs}$, the electron density distribution (Fig.~\ref{fig1}(c)) indicates that a C--H bond has broken, liberating a free proton. The cleavage of C--H bonds and the generation of free protons are mutually verified by the statistical results of \(N_{\mathrm{C-H}}\) and $N_{\mathrm{H^+}}$ shown in Figs.~\ref{fig4}(a) and \ref{fig4}(b). The Coulomb potential of the $\ch{H^+}$ localizes surrounding electrons (yellow region in Fig.~\ref{fig1}(c)), binding them in the vicinity of the proton. This localization significantly reduces the density of quasi-free electrons, thereby decreasing $|\Delta E(\tau)/E|$ and enhancing the transmitted THz electric field. For $\tau > \SI{350}{fs}$, the electron-capturing effect of the protons saturates, and the quasi-free electron density resumes its increase, causing $|\Delta E(\tau)/E|$ to return to its rising trend.

Figure~\ref{fig4}(d) presents the theoretically simulated change in THz absorption, $\Delta E$, of the ethanol system as a function of $\tau$ under different pump intensities. The simulations reveal a rapid decrease followed by a rapid increase in $\Delta E$ within the delay range of \SIrange{150}{350}{fs} for pump intensities investigated. This behavior is in good agreement with the experimental results shown in Fig.~\ref{fig2}(c).

Meanwhile, we repeated the OPTP experiment using ethyl acetate,  \ch{CH3COOC2H5}, which contains C--H bonds but lacks 
O--H bonds. 
In the present strong-field ionization experiment of \ch{CH3COOC2H5},
the evolution of $|\Delta E(\tau)/E|$ with $\tau$ exhibits a behavior similar to that of the 
\ch{C2H5OH} system, with a distinct dip also observed near $\tau \approx 300$~fs, as highlighted by the pink curve in Fig.~\ref{fig2}(c).
This confirms that the free protons originate from the C--H bonds in the alkyl group rather than from the O--H bonds in the hydroxyl group.

To provide further evidence for the C--H origin of this dip, we performed the same OPTP measurements on deuterated ethanol variants. As shown in Supplement Materials, the $|\Delta E/E|$ for pure hydroxyl-deuterated ethanol (\ch{C2H5OD}) at $I = \SI{6.3e13}{\watt\per\centi\meter\squared}$ exhibits a V-shaped dip that is essentially identical to that of pure ethanol, confirming that selective deuteration of the O--D group does not significantly affect the transient quasi-free electron response. In contrast, fully deuterated ethanol (\ch{C2D5OD}), the dip becomes even clearly deeper and has a much slower recovery progress, indicating a stronger and longer-lasting suppression of the free-electron density.  
This may arise from a slightly stronger and more structurally ordered hydrogen-bonding network in \ch{C2D5OD} than in \ch{C2H5OH}, originating from nuclear quantum effects and the lower zero-point vibrational energy, analogous to the case of \ch{D2O} versus \ch{H2O}\cite{Giubertoni2023jpcb,Ceriotti2016,liang2023jpcl}.
Together with the \ch{CH3COOC2H5} results, these isotope substitution experiments provide strong evidence that the dip in the THz transient absorption originates from free protons released specifically from C--H bonds in the ethyl group.
Most previous studies have focused on O--H bond breaking because the hydroxyl proton can rapidly transfer through the hydrogen-bond network. Our data also show a distinct change in the OPTP slope at $\tau \approx 100$--$200$ fs 
in both ethanol and pure water, as shown 
in Fig.~\ref{fig2}(c) and Fig.~\ref{fig2}(e), respectively, but is absent in \ch{CH3COOC2H5}. This feature might be related to O–H bond breaking and proton transfer,
such ultrafast proton transfer between neighboring water molecules following ionization has been directly observed in liquid water \cite{Loh2020}.
In the present system, the proton may be rapidly transferred through the hydrogen-bond network rather than remaining in the local environment. 
A single ethanol molecule contains five C--H bonds but only one O--H bond. The statistical predominance of C--H over O--H suggests that the C--H cleavage channel is intrinsically more productive in generating free protons than the O--H pathway. Consequently, the V-shaped dip induced by C–H cleavage is considerably more pronounced than the slope change caused by O–H bond breaking.

Rt-TDDFT not only provides predictions for the bond statistics and THz absorption of pure ethanol solutions after photoionization but also offers insights into the behavior of ethanol-water mixture
subjected to strong-field photoionization. 
Compared to the pure ethanol case, the average value of \(N_{\mathrm{C-H}}\) for \ch{C2H5OH}--\ch{H2O} is significantly larger than that for \ch{C2H5OH}, while the average value of $N_{\mathrm{H^+}}$ for the mixture is notably smaller as shown in Figs.~\ref{fig4}(a) and \ref{fig4}(b). This indicates that the probability of C--H bond cleavage and the number of free protons formed are lower in the ethanol-water mixture than in pure ethanol. The underlying reasons are illustrated in Figs.~\ref{fig4}(e) and (f), where the statistics of the number of O--H bonds, $N_{\mathrm{O-H}}$, are shown within the liquid unit cell for both systems. In the \ch{C2H5OH}--\ch{H2O}, both C--H bond cleavage (from ethanol) and O--H bond cleavage (from water) occur. However, protons generated from either cleavage pathway can recombine with oxygen atoms to form new O--H bonds. Consequently, at $\tau = \SI{500}{fs}$, an increase in the number of O--H bonds is observed in Figs.~\ref{fig4}(e) and (f). Thus, theory predicts that in the \ch{C2H5OH}--\ch{H2O} mixture, the number of free protons is very low, and the effect of free protons trapping electrons is absent; hence, the electron density should not exhibit the abrupt change observed in the pure ethanol.

In summary, using optical-pump terahertz-probe (OPTP) spectroscopy, we observe a sub-100-fs V-shaped modulation of quasi-free electron density during the photoionization of liquid \ch{C2H5OH} and \ch{CH3COOC2H5}. Rt-TDDFT simulations attribute this V-shaped dip to electron localization around free protons released via C–H bond cleavage. Furthermore, experimental and theoretical results reveal that trace amounts of water reinforce the hydrogen-bonded network in ethanol solutions, thereby facilitating proton transfer to form new O–H bonds. Our findings demonstrate that photoionization can directly generate free protons from typically inert C–H bonds in liquid environments. These free protons persist only on a timescale of a few hundred femtoseconds before being scavenged by surrounding molecules via the hydrogen-bonded network. This work provides critical insights into ultrafast C–H activation dynamics in condensed phases.

\medskip
\begin{acknowledgments}
This work is supported by the National Key Research and Development Program of China (Grant No.~2022YFA1604302), the National Natural Science Foundation of China (NSFC) (Grants No.~12334011, No.~12541505, No.~12374262, No.~12674347, and No.~12450401), the Chinese Academy of Sciences (CAS) (Grant No.~YSBR-047), and the Natural Science Foundation of Henan (Grant No. 252300421304). We acknowledge the computational resource of the HPC platform in ShanghaiTech University.
\end{acknowledgments}

\bibliographystyle{apsrev4-2} 
\bibliography{main}
\end{document}


\renewcommand{\theequation}{S\arabic{equation}}
\renewcommand{\thefigure}{S\arabic{figure}}
\renewcommand{\thetable}{S\arabic{table}}

\title{Supplementary Material: Ultrafast Proton-Induced Collective Dynamics Regulated by Water-Mediated Hydrogen-Bond Networks in Liquids}

\author{Xiaokun Liu}
\thanks{These authors contributed equally to this work.}
\affiliation{Center for Transformative Science and School of Physical Science and Technology, ShanghaiTech University, Shanghai 201210, China}
\author{Bozhu Chen}
\thanks{These authors contributed equally to this work.}
\affiliation{Songshan Lake Materials Laboratory, Dongguan 523808, China}
\author{Yizhu Zhang}
\email{zhangyizhu@tiangong.edu.cn}
\affiliation{School of Electrical Engineering, Tiangong University, Tianjin 300387, China}
\author{Jiyu Xu}
\email{jiyuxu@iphy.ac.cn}
\affiliation{Beijing National Laboratory for Condensed Matter Physics, Institute of Physics, Chinese Academy of Sciences, Beijing 100190, China}
\author{Meng Li}
\affiliation{Center for Transformative Science and School of Physical Science and Technology, ShanghaiTech University, Shanghai 201210, China}
\affiliation{Shanghai Advanced Research Institute, Chinese Academy of Sciences, Shanghai 201210, China}
\author{Yanjun Gao}
\affiliation{Shanghai Advanced Research Institute, Chinese Academy of Sciences, Shanghai 201210, China}
\author{Ji Qi}
\affiliation{School of Physics, Henan Normal University, Xinxiang 453007, China}
\author{Yuqian Yang}
\affiliation{Center for Transformative Science and School of Physical Science and Technology, ShanghaiTech University, Shanghai 201210, China}
\author{Wennan Wang}
\affiliation{Center for Transformative Science and School of Physical Science and Technology, ShanghaiTech University, Shanghai 201210, China}
\author{Sheng Meng}
\email{smeng@iphy.ac.cn}
\affiliation{Beijing National Laboratory for Condensed Matter Physics, Institute of Physics, Chinese Academy of Sciences, Beijing 100190, China}
\affiliation{Songshan Lake Materials Laboratory, Dongguan 523808, China}
\affiliation{School of Physical Sciences, University of Chinese Academy of Sciences, Beijing 100190, China}
\author{Yuhai Jiang}
\email{jiangyh3@shanghaitech.edu.cn}
\affiliation{Center for Transformative Science and School of Physical Science and Technology, ShanghaiTech University, Shanghai 201210, China}
\affiliation{School of Physics, Henan Normal University, Xinxiang 453007, China}

\date{\today}

\maketitle

\section{Experimental Methods}
\subsection{Optical Pump–Terahertz Probe Setup}

The optical pump–terahertz probe (OPTP) setup~\cite{li2025apl}was driven by a Ti:sapphire regenerative amplifier (Spectra Physics) delivering pulses at a central wavelength of \SI{800}{nm}, with a pulse duration of \SI{60}{fs}, a repetition rate of \SI{1}{kHz}, and a pulse energy of \SI{2.4}{\milli\joule}. The polarization was horizontal (p-polarized). The laser output was split by a beam splitter: the reflected part (approximately 30\%, \SI{0.7}{\milli\joule}) served as the pump beam, while the transmitted part (approximately 70\%, \SI{1.7}{\milli\joule}) was used for THz generation and detection. The transmitted beam was further divided, with approximately 99\% directed to THz generation and the remaining approximately 1\% used for electro-optic sampling of the THz waveform.

\subsection{Pump and THz Beam Characterization}
The pump beam was attenuated by a variable neutral density filter to achieve pulse energies between \SI{70}{\micro\joule} and \SI{200}{\micro\joule}. The beam passed through a motorized delay stage with a maximum travel of \SI{150}{\milli\meter} and a minimum step size of \SI{1}{\micro\meter}, allowing a pump–probe delay range from \SI{-150}{ps} to \SI{950}{ps} with a minimum delay increment of \SI{6.7}{fs}. The beam diameter was reduced from ~\SI{12.3}{\milli\meter} to ~\SI{5}{\milli\meter} by a telescope consisting of lenses with focal lengths \SI{+150}{\milli\meter} and \SI{-50}{\milli\meter}. It was then focused onto the sample by a lens with a focal length of \SI{250}{\milli\meter}, producing a spot size of approximately ~\SI{50}{\micro\meter} in diameter at the focus. The peak intensity at the focus was adjustable in the range \SIrange{3e13}{11e13}{\watt\per\centi\meter\squared}. After passing through the sample, the residual pump light was blocked by a Teflon (PTFE) film.

The beam for THz generation (centered at \SI{800}{nm}) was frequency-doubled in a \SI{200}{\micro\meter}-thick $\beta$-BBO crystal to produce the second harmonic (\SI{400}{nm}) with a conversion efficiency of ~30\%\cite{Zhao2024cp,Zhang2020ol,Zhao2025apl,Zhang2022ol}. The second harmonic was vertically polarized (type-I phase matching). A dual-wavelength wave plate placed after the BBO
crystal acted as a half-wave plate at \SI{800}{nm}
and a full-wave plate at \SI{400}{nm}. It adjusted
the relative polarization angle between the
fundamental and second harmonic to maximize
the THz yield. The fundamental and second harmonic were focused together by a lens to form a plasma in air, which emits broadband THz radiation. The remaining optical beams were blocked by a PTFE filter. The generated THz beam was collected by an off-axis parabolic mirror with a focal length of \SI{101.4}{\milli\meter} and then focused onto the sample by another parabolic mirrors. The THz spot size at the sample was approximately \SI{300}{\micro\meter} in diameter. The measured average power of the THz beam was approximately
\SI{60}{\micro\watt}.

After the sample, the THz beam was recollimated and focused onto a \SI{0.5}{\milli\meter}-thick, (110)-cut ZnTe crystal for electro-optic sampling. A small portion of the \SI{800}{nm} beam (probe) was focused onto the same ZnTe crystal. The THz field induces birefringence in the crystal, which was detected by a quarter-wave plate, a Wollaston prism, and a pair of balanced photodiodes connected to a lock-in amplifier. The effective detection bandwidth of the electro-optic
sampling system was approximately
\SIrange{0.1}{3}{\tera\hertz}.
 The pump beam was mechanically chopped, and the lock-in amplifier detected the pump-induced modulation of the THz waveform. The entire setup was enclosed in a laboratory environment where the humidity was maintained at about 32\% to reduce water vapor absorption. All measurements were performed at a room temperature of \SI{18}{\degreeCelsius}.

\subsection{Sample Preparation and Characterization}

Ethanol (\ch{C2H5OH}, 99.7\% purity, Macklin, catalog no.~E809061), hydroxyl-deuterated ethanol (\ch{C2H5OD}, J\&K Scientific, catalog no.~911345, 99 atom\% D), fully deuterated ethanol (\ch{C2D5OD}, Cambridge Isotope Laboratories, catalog no.~DLM-31-10X1, 99\% D), ethyl acetate (\ch{CH3COOC2H5}, 99.5\% purity, Xilong Scientific, catalog no.~1300020101600), heavy water (\ch{D2O}, Leyan, catalog no.~C400004, 99.8\% D), and distilled water (\ch{H2O}) were used to prepare the samples. Samples were prepared by mixing the appropriate ethanol isotopologue with \ch{H2O} or \ch{D2O} in a volumetric flask. The volume ratios were calculated from the target molar ratios using the densities of the respective components. Samples were handled rapidly under ambient conditions and measured immediately after preparation to limit exposure to atmospheric moisture.


A gravity-driven, wire-guided flow jet was used to generate a stable, free-standing liquid film. 
The jet had no windows, which avoided transmission artifacts.  Continuous liquid flow
helped reduce thermal accumulation in the
excitation volume. The setup consisted of an upper reservoir equipped with a nozzle, a 3D-printed film-forming frame, a lower reservoir, a peristaltic pump, and a pulse damper \cite{Wang2014,Tan2021}. Liquid flowed from the upper reservoir through a flexible tube with an inner diameter of \SI{2}{\milli\meter} and an outer diameter of \SI{4}{\milli\meter} into the film-forming frame. After passing through a \SI{5}{\milli\meter} horizontal buffer region, the liquid was guided by two parallel tungsten wires, each \SI{40}{\micro\meter} in diameter and separated by \SI{5}{\milli\meter}, to form a thin film. The liquid then passed through a conical structure at the bottom of the frame and was collected as droplets in the lower reservoir. A peristaltic pump (BT00L with a YZ15 pump head) recirculated the liquid through a pulse damper to minimize flow pulsations. The film-forming frame was mounted on a KM200 kinematic mount for precise alignment. 
The jet thickness was approximately \SI{50}{\micro\meter}, as estimated from THz time-domain transmission measurements. Neglecting interfacial reflection losses and multiple internal reflections, the transmitted and incident electric-field amplitudes are related by
\[
    \frac{I_{\mathrm{film}}(\nu)}{I_{\mathrm{ref}}(\nu)}
    =
    \left|
    \frac{E_{\mathrm{film}}(\nu)}{E_{\mathrm{ref}}(\nu)}
    \right|^{2}
    =
    \exp[-\alpha(\nu)d]
\]
where $\alpha(\nu)$ is the power absorption coefficient and $d$ is the film thickness. The film thickness was estimated from the measured THz field transmission at \SI{1}{\tera\hertz},
using literature-reported power absorption coefficients of approximately \SI{220}{\per\centi\meter} for pure water and \SI{60}{\per\centi\meter} for pure ethanol at this frequency~\cite{Kindt1996JPC}.

\subsection{Data Analysis and Signal Extraction}

The OPTP measurements involved two time coordinates:
the THz sampling time $t$, which resolves the transmitted
THz waveform, and the pump--probe delay $\tau$, which
tracks the pump-induced dynamics. The differential THz
field was defined as
\begin{equation}
    \Delta E(t,\tau)
    =
    E_{\mathrm{pump-on}}(t,\tau)
    -
    E_{\mathrm{pump-off}}(t).
\end{equation}

The unpumped reference waveform
$E_{\mathrm{pump-off}}(t)$ was acquired separately
with the pump beam blocked and the THz-generation
beam mechanically chopped.
The transient traces were recorded as a function of
pump--probe delay $\tau$ at a fixed THz sampling
time $t=t_p$, corresponding to the peak of the
unpumped reference waveform. Here, $t_p=0$~ps
corresponds to the THz sampling time used to extract
the transient traces in Fig.~2 of the main text.
The reference field was defined as
$E(t_p)=E_{\mathrm{pump-off}}(t_p)$, and the normalized
signal was calculated as $\Delta E(t_p,\tau)/E(t_p)$, abbreviated as $\Delta E/E$
in the main text.
The magnitude $\Delta E(t_p,\tau)/E(t_p)$ was used to
display the transient dynamics in the figures. The signed signal $\Delta E(t_p,\tau)/E(t_p)$ retained the same sign throughout the dip.

The pump-induced change in the transmitted THz field
provides a probe of the transient conductivity response.
In a simple transport picture, the conductivity is
given by $\sigma=ne\mu$, where $n$ is the carrier
number density, $e$ is the elementary charge, and
$\mu$ is the carrier mobility. The THz response also
depends on frequency-dependent scattering and localization
dynamics. Therefore, the measured field change alone
does not distinguish changes in carrier density from
changes in mobility. Its microscopic interpretation
is considered together with the rt-TDDFT simulations.




\subsection{OPTP Results for Pure Hydroxyl-Deuterated
Ethanol and Fully Deuterated Ethanol}

The isotope-substitution measurements were performed in
separate experimental sessions. For each isotope-substituted
sample, a pure ethanol reference was measured immediately
beforehand in the same session under matched measurement
conditions. Each isotope-substituted trace was compared
with its corresponding ethanol reference.
These measurements were performed at a pump intensity of
\SI{6.3e13}{\watt\per\centi\meter\squared}.

\begin{figure}[htbp]
    \centering
    \includegraphics[width=\linewidth]{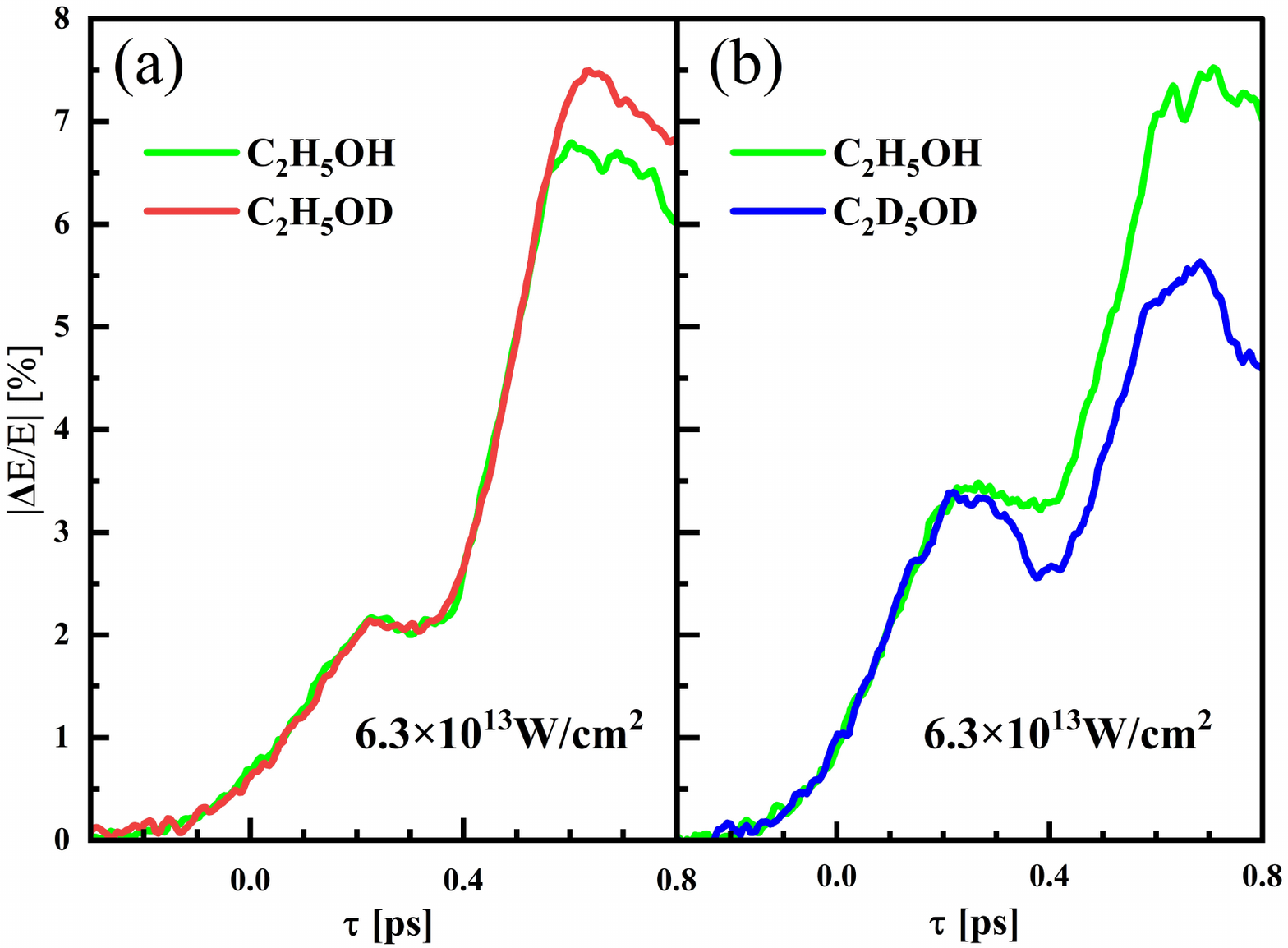} 
    \caption{Normalized differential THz-field dynamics
    $|\Delta E/E|$ for
    (a) hydroxyl-deuterated ethanol (\ch{C2H5OD}) and
    (b) fully deuterated ethanol (\ch{C2D5OD}), compared
    with their corresponding pure ethanol references
    (green curves). Measurements were performed at a
    pump intensity of
    \SI{6.3e13}{\watt\per\centi\meter\squared}.
    }\label{figs1}
\end{figure}

As shown in Fig.~\ref{figs1}(a), the
$|\Delta E/E|$ dynamics of pure \ch{C2H5OD}
exhibit a V-shaped dip similar to that of pure ethanol,
indicating that hydroxyl deuteration has little effect
on the measured transient THz response. In contrast,
fully deuterated ethanol (\ch{C2D5OD}) exhibits a
deeper dip and a slower recovery
[Fig.~\ref{figs1}(b)]. The contrast between
hydroxyl-only and full deuteration supports the
involvement of the ethyl hydrogen sites in the
dynamics responsible for the THz dip.

\section{Computational details}

\subsection{Molecular Dynamics Simulations}

We employ real-time time-dependent density functional theory (rt-TDDFT) simulations to investigate the photoinduced ultrafast dynamics of ethanol and an ethanol--water mixture. The rt-TDDFT molecular dynamics (MD) simulations based on Ehrenfest dynamics were performed via the Time Dependent Ab-initio Package (TDAP)~\cite{Lian2020} implemented in the Quantum ESPRESSO suite of programs~\cite{Giannozzi2009}. We adopt the optimized norm-conserving Vanderbilt (ONCV) pseudopotentials~\cite{Hamann2013}, and the energy cutoffs for wavefunction and charge density were set to be 50 Ry and 400 Ry for all MD simulations, respectively. We employ the optB88-vdW~\cite{Klimes2010} method to describe the ethanol and ethanol--water mixture. An $11~\text{\AA} \times 11~\text{\AA} \times 11~\text{\AA}$ cubic cell with 14 ethanol molecules was used for ethanol solution MD simulations, while 3 water molecules were additionally added to the same cubic cell to construct the ethanol--water mixture. The Brillouin zone was sampled with a $1 \times 1 \times 1$ Monkhorst-Pack k-point mesh. The unoccupied electronic states considered in the MD simulations were 60 and 68 for ethanol and ethanol--water mixture, respectively. The time-dependent Kohn-Sham wave functions based on rt-TDDFT evolve in NVE ensemble using Verlet algorithm, with a time step of \SI{0.04}{fs} for nuclei and \SI{0.08}{as} for electrons. 

Before applying the laser field, the solution was first equilibrated for 5 ps at 300 K in the NVT ensemble in ab initio molecular dynamics (AIMD) simulation~\cite{Giannozzi2009}, and last representative crystal configurations were extracted for laser irradiation with a Gaussian envelope, i.e.,
\( E(t) = E_0 \exp\left[-\frac{(t - t_0)^2}{2\sigma^2}\right] \cos\left(\frac{2\pi c}{\lambda} t\right) \)
where $c$ was the speed of light and $\lambda$ was the wavelength. We used a laser pulse with $\lambda = 800$ nm, a width of $\sigma = 4$ fs centered at $t_0 = 20$ fs. The maximum electric field $E_0$ was \SI{2.0}{\volt\per\angstrom} or \SI{2.2}{\volt\per\angstrom} for the nonadiabatic simulations, which correspond to the pump intensity of \SI{5.31e13}{\watt\per\centi\meter\squared} and \SI{6.42e13}{\watt\per\centi\meter\squared}, respectively. Periodic electric fields were guaranteed through velocity gauge in rt-TDDFT simulations. 

\subsection{Terahertz Transmissivity Calculation}

We calculate the electronic terahertz transmissivity based on the electronic conductivity, which can be deduced from Ohm's law. The corresponding basic equations were listed below\cite{Philipp2026}:

\begin{align}
J(\omega) &= \sigma E(\omega) \label{eq:Jsigma} \\
\sigma &= \sigma_1 + i\sigma_2 \label{eq:sigma} \\
\varepsilon &= \varepsilon_1 + i\varepsilon_2 \label{eq:eps} \\
\varepsilon_1 &= 1 - \frac{4\pi\sigma_2}{\omega} \label{eq:eps1} \\
\varepsilon_2 &= \frac{4\pi\sigma_1}{\omega} \label{eq:eps2} \\
n &= \sqrt{\frac{\sqrt{\varepsilon_1^2 + \varepsilon_2^2} + \varepsilon_1}{2}} \label{eq:n} \\
k &= \sqrt{\frac{\sqrt{\varepsilon_1^2 + \varepsilon_2^2} - \varepsilon_1}{2}} \label{eq:k} \\
T &= \frac{4n}{(1+n)^2 + k^2} \label{eq:T}
\end{align}

The $J(\omega)$ and $E(\omega)$ in Eq.~\eqref{eq:Jsigma} were the current density and electric field in the frequency domain. The $\sigma(\omega)$ and $\varepsilon(\omega)$ were electronic conductivity and dielectric coefficient, respectively, which were complex in nature. The $n$, $k$ and $T$ were refractive index, extinction coefficient and electronic transmissivity, respectively. 

After the rt-TDDFT MD simulations with 800 nm laser pulse, we can calculate the electronic terahertz transmissivity at each time step based on Eqs.~\eqref{eq:Jsigma}--\eqref{eq:T}. The frequency-domain terahertz photocurrent needed in Eq.~\eqref{eq:Jsigma} can be calculated via Fourier transforming the time-domain terahertz photocurrent, which can be obtained by running another rt-TDDFT MD simulation with a terahertz field applied to the atomic configuration at each time step of 800 nm rt-TDDFT trajectory. However, this was very time consuming. To save computational resources, we calculate the frequency-domain terahertz transmissivity spectrum for some particular moments in the 800 nm rt-TDDFT trajectory, i.e., 0 fs, 100 fs, 200 fs, 250 fs, 350 fs, 450 fs, and the continuous temporal transmissivity curves were obtained via interpolation. Due to the limited time scale of the simulations, the frequency range of the simulated terahertz field was 0--16 THz, but the transmissivity was calculated within 0--3 THz, consistent with the experimental measurement. 
\medskip
\bibliographystyle{apsrev4-2}
\bibliography{main}